\documentclass[aip,jcp,reprint,amsmath,amssymb,floatfix,citeautoscript,nofootinbib]{revtex4-1}
\usepackage{times}
\usepackage{mathptmx}
\DeclareMathAlphabet{\mathcal}{OMS}{cmsy}{m}{n} 
\usepackage[T1]{fontenc}
\usepackage{algorithm}
\usepackage{algpseudocode}
\usepackage{bm}
\usepackage{amsmath}
\usepackage{amssymb}
\usepackage{xcolor}
\usepackage{graphicx}
\usepackage{hyperref}
\hypersetup{
    colorlinks=true,
    linkcolor=blue,
    urlcolor=blue,
    citecolor=blue,
}
\usepackage{cleveref}
\usepackage{braket}
\usepackage[font={small}]{caption}
\usepackage{mhchem}
\usepackage{multirow}

\newcommand{\mi}{\mathrm{i}}
\newcommand{\me}{\mathrm{e}}
\newcommand{\mat}[1]{\mathbf{#1}}

\definecolor{myred}{rgb}{0.8,0,0}

\begin{document}

    \title{Periodic Local Coupled-Cluster Theory for Insulators and Metals}

    \author{Hong-Zhou Ye}
    \email{hzyechem@gmail.com}
    \affiliation{Department of Chemistry, Columbia University, New York, NY, 10027, USA}
    \author{Timothy C. Berkelbach}
    \email{tim.berkelbach@gmail.com}
    \affiliation{Department of Chemistry, Columbia University, New York, NY, 10027, USA}
    \affiliation{Initiative for Computational Catalysis, Flatiron Institute, New York, NY, 10010, USA}

    \begin{abstract}
    We describe the implementation details of periodic local coupled-cluster theory with single and double excitations (CCSD) and perturbative triple excitations [CCSD(T)] using local natural orbitals (LNOs) and $k$-point symmetry.
    We discuss and compare several choices for orbital localization, fragmentation, and LNO construction.
    By studying diamond and lithium, we demonstrate that periodic LNO-CC theory can be applied with equal success to both insulators and metals, achieving speedups of two to three orders of magnitude even for moderately sized $k$-point meshes.
    Our final predictions of the equilibrium cohesive energy, lattice constant, and bulk modulus for diamond and lithium are in good agreement with previous theoretical predictions and experimental results.
    \end{abstract}

    \maketitle

    \section{Introduction}

    Correlated wavefunction methods, which describe correlation at the level of the many-electorn wavefunction, have attracted increasing attention in computational materials science~\cite{Pisani12PCCP,Booth13Nature,Yang14Science,McClain17JCTC,Gruber18PRX,Zhang19FM,Lau21JPCL,Kubas16JPCL,Sauer19ACR,Brandenburg19JPCL,Maristella19JCTC,Schafer21JCPb,Cui22Science,Mullan22JCP,Shi23JACS}.
    Among these methods, coupled-cluster (CC) theory~\cite{Crawford00Inbook,Bartlett2007} is particularly notable for its size extensivity and its ability to achieve high accuracy even when truncated at low orders.
    The most commonly used variants are CCSD, which includes single and double excitations, and CCSD(T)~\cite{Raghavachari89CPL}, which adds a perturbative treatment of triple excitations.
    The latter is often referred to as the gold standard of molecular quantum chemistry due to its ability to deliver chemical accuracy, typically defined as within $1$~kcal/mol, across a wide range of weakly correlated systems~\cite{Tajti04JCP,Curtiss07JCP,Bartlett2007}.
    Recently, CCSD and CCSD(T) have been applied to simple materials with periodic boundary conditions, demonstrating good agreement with experimental results for insulators and semiconductors~\cite{Booth13Nature,McClain17JCTC,Gruber18PRX}, metals~\cite{Mihm21NCS,Neufeld22JPCL,Neufeld23PRL,Masios23PRL}, molecular crystals~\cite{Yang14Science}, and liquid water~\cite{Yu23JPCL,Chen23JCTC}.
    However, the high computational cost of these CC methods, which scale as the sixth or seventh power of the system size, significantly limits their routine application in materials studies.

    In molecular quantum chemistry, reduced-scaling techniques have been developed over the past few decades to expand the size of systems that can be studied with CC theory.
    These methods can be roughly categorized into two approaches.
    One approach is based on fragment-based methods, which operate in a divide-and-conquer manner.
    In this strategy, a large molecule is divided into local fragments, each of a smaller size that can be solved by canonical CC.
    Examples of this approach include the divide-expand-consolidate method~\cite{Ziolkowski10JCP,Kristensen12PCCP,Kjargaard17JCP,Barnes19JPCA}, the cluster-in-molecule method~\cite{Li02JCC,Li04JCP,Li06JCP,Li09JCP,Wang19JCTC,Wang22JCTC} and its local natural orbital (LNO) implementation~\cite{Rolik11JCP,Rolik13JCP,Nagy17JCP,Nagy19JCTC,GyeviNagy21JCTC,Szabo23JCTC}, as well as many recently developed quantum embedding methods~\cite{Cui20JCTC,Ye20JCTC,Nusspickel22PRX,Shee24arXiv}.
    The other approach focuses on seeking sparse or low-rank representations of the CC wavefunction, which leads to a reduced-scaling solution of the original, fully-coupled CC amplitude equations.
    This can be achieved either in a localized occupied orbital basis by compressing the virtual space, as seen in pair natural orbital-based methods~\cite{Neese09JCPa,Neese09JCPb,Hansen11JCP,Huntington12JCP} or in a particle-hole basis by factorizing the CC amplitudes globally, as in the rank-reduced CC method~\cite{Parrish19JCP,Hohenstein19JCP,Hohenstein22JCP}.
    Both approaches have been combined with existing low-rank factorization methods for the electron-repulsion integrals, such as local density fitting~\cite{Werner03JCP,Schutz04JCP,Tew18JCP} and tensor hypercontraction~\cite{Hohenstein12JCP,Parrish12JCP,Parrish13PRL}, to enable efficient CC calculations of molecules.

    Recently, we extended molecular LNO-based CC (henceforth referred to as LNO-CC) to periodic systems with $\Gamma$-point Brillouin zone sampling and presented preliminary applications to adsorption, dissociation~\cite{Ye23arXiv}, and vibrational spectroscopy~\cite{Ye24FD} of small molecules on surfaces at the CCSD(T) level.
    In this work, we provide full details of a more generic framework for periodic local CC theory, including the necessary extensions to exploit lattice translation symmetry with $k$-point sampling, and we compare different methods for defining fragments and constructing LNOs.
    In \cref{sec:results_and_discussion}, we benchmark the accuracy and computational efficiency of LNO-CC by calculating the ground-state thermochemical properties, including the equilibrium lattice constant, bulk modulus, and cohesive energy, of two crystals: diamond and body-centered cubic (BCC) lithium.
    In both cases, LNO-CC quickly converges all properties to the canonical CC limit while keeping only a small fraction of LNOs within the active space, resulting in more than $100$-fold speedup over the canonical CCSD with $k$-points, even for a medium-sized $k$-point mesh of $3\times3\times3$.
    Our final numbers in the thermodynamic limit (TDL) and the complete basis set (CBS) limit obtained at the CCSD(T) level for diamond and the CCSD level for lithium exhibit very good agreement with both previous calculations and available experimental data.

    \section{Theory}
    \label{sec:theory}

    A spin-restricted HF reference wavefunction that preserves the lattice translational symmetry is assumed throughout this work.
    The Brillouin zone is sampled with a uniform $k$-point mesh of size $N_k$.
    The HF reference can be equivalently represented in both a unit cell with $N_k$ $k$-points and a supercell containing $N_k$ unit cells with a single $k$-point.
    In the former, the orbitals are denoted as $\psi_{p\bm{k}}(\bm{r})$ with orbital energies $\varepsilon_{p\bm{k}}$, where $p = 1, 2, \ldots, n_{\textrm{MO}}$.
    In the latter, the orbitals are denoted as $\psi_{p}(\bm{r})$ with orbital energies $\varepsilon_{p}$, where $p = 1, 2, \ldots, N_{\textrm{MO}}$ with $N_{\textrm{MO}} = N_k n_{\textrm{MO}}$.
    A tensor expressed in one representation can always be converted into the other through a unitary transform
    \begin{equation}
        A_{p\cdots}
            = \sum_{\bm{k}q} \braket{\psi_{p} | \psi_{q\bm{k}}} A_{q\bm{k}\cdots}
    \end{equation}
    This equivalence between the two representations will be exploited frequently in the following sections.
    The context should make it clear whether a tensor is in the $k$-point orbital basis or the supercell orbital basis.
    In both representations, we use $i,j,k,\cdots$ to index occupied orbitals, $a,b,c,\cdots$ for virtual orbitals, and $p,q,r,\cdots$ for unspecified orbitals.
    The electronic Hamiltonian in the supercell HF orbital basis reads
    \begin{equation}
        \hat{H}
            = \sum_{\sigma} \sum_{pq} h_{pq} c^{\dagger}_{p\sigma} c_{q\sigma}
            + \frac{1}{2} \sum_{\sigma\sigma'} \sum_{pqrs} V_{pqrs}
            c^{\dagger}_{p\sigma} c^{\dagger}_{r\sigma'} c_{s\sigma'} c_{q\sigma}
    \end{equation}
    where $\sigma$ labels spin and the electron-repulsion integrals (ERIs) $V_{pqrs}$ are in $(11|22)$ notation.

    \subsection{Localized orbitals and local fragments}
    \label{subsec:localized_orbitals}

    We construct localized orbitals (LOs) within the supercell directly from the $k$-point HF orbitals
    \begin{equation}    \label{eq:lo_from_kpt_hf}
        \phi_{\alpha\bm{R}}(\bm{r})
            = \sum_{\bm{k}}^{N_k} \me^{\mi\bm{k}\cdot\bm{R}}
            \sum_{p} \psi_{p\bm{k}}(\bm{r}) U_{p\alpha}(\bm{k})
    \end{equation}
    where $\bm{R}$ labels the $N_k$ unit cells within the supercell, and there are $n_{\textrm{LO}}$ LOs within each cell, giving rise to a total of $N_{\textrm{LO}} = N_k n_{\textrm{LO}}$ LOs.
    These LOs preserve the lattice translational symmetry
    \begin{equation}    \label{eq:lo_translation_symm}
        \phi_{\alpha\bm{R}}(\bm{r})
            = \phi_{\alpha\bm{0}}(\bm{r} - \bm{R})
    \end{equation}
    meaning that only LOs within a reference unit cell, chosen here to be $\bm{R} = \bm{0}$, need be determined explicitly.
    From now on, we drop the cell label $\bm{R}$ and let $\phi_{\alpha} \equiv \phi_{\alpha\bm{0}}$ denote the LOs within the reference cell.
    The unitary rotations $U_{p\alpha}(\bm{k})$ in \cref{eq:lo_from_kpt_hf} are variational degrees of freedom chosen to minimize the spatial extent of the LOs.
    As will be explained in \cref{subsec:periodic_lno_cc}, LNO-CC requires the LOs to span the occupied space
    \begin{equation}    \label{eq:lo_span_occ}
        \sum_{\alpha} U_{i\alpha}(\bm{k}) U^*_{j\alpha}(\bm{k})
            = \delta_{ij}
    \end{equation}
    for all $\bm{k}$ in the $k$-point mesh.
    \Cref{eq:lo_span_occ} is a condition satisfied by many commonly used LO types~\cite{Boys60RMP,Pipek89JCP,Marzari12RMP,Jonsson17JCTC,Knizia13JCTC}.
    In this work, we will focus on two types of LOs: Pipek-Mezey (PM) orbitals~\cite{Pipek89JCP,Jonsson17JCTC}, which are similar to Boys orbitals (more commonly known as maximally localized Wannier functions in materials science), and intrinsic atomic orbitals~\cite{Knizia13JCTC} (IAOs).
    For PM orbitals, the summation over $p$ in \cref{eq:lo_from_kpt_hf} is restricted to occupied orbitals, producing $n_{\textrm{LO}} = n_{\textrm{occ}}$ LOs that span exactly the occupied space.
    In contrast, IAOs span both the occupied space and the valence part of the virtual space~\cite{Knizia13JCTC}, making $n_{\textrm{LO}}$ slightly greater than $n_{\textrm{occ}}$ but still much smaller than the total number of orbitals $n_{\textrm{MO}}$.
    For systems where the occupied space can be readily localized, both PM orbitals and IAOs can be used in a LNO-CC calculation.
    In cases where localization of the occupied orbitals is not feasible, IAOs are the preferred option.
    We will examine examples from both categories in \cref{sec:results_and_discussion}.
    In what follows, it is convenient to use the supercell representation of the LO coefficient matrix, denoted as $U_{p\alpha}$, which has dimensions $N_{\textrm{MO}} \times n_{\textrm{LO}}$.

    For generality, we introduce $n_{\textrm{frag}}$ local fragments by partitioning the $n_{\textrm{LO}}$ LOs within the reference cell into disjoint subsets such that
    \begin{equation}
        n_{\textrm{LO}}
            = \sum_{F}^{n_{\textrm{frag}}} n_{\textrm{LO}}^{(F)}
    \end{equation}
    where $n_{\textrm{LO}}^{(F)}$ is the number of LOs in fragment $F$.
    For PM orbitals, which are not necessarily localized on a single atom, this partitioning is best done at the individual LO level, i.e.,~$n_{\textrm{LO}}^{(F)} = 1$ for all fragments, producing $n_{\textrm{frag}} = n_{\textrm{occ}}$ fragments.
    This is also the scheme adopted by the original molecular LNO-CC method~\cite{Rolik11JCP}.
    For IAOs, using individual LOs as fragments breaks the rotational invariance of the LNO-CC correlation energy, and thus we group all IAOs localized on an atom as a fragment.
    This leads to $n_{\textrm{frag}} = n_{\textrm{atom}}$ fragments, where $n_{\textrm{atom}}$ is the number of atoms within the reference cell.
    We emphasize that the number of fragments $n_{\textrm{frag}}$ does not increase with $N_k$, due to the equivalence of fragments in other cells by lattice translation symmetry.

    \subsection{Periodic LNO-CCSD and LNO-CCSD(T)}
    \label{subsec:periodic_lno_cc}

    Periodic LNO-CCSD approximates periodic canonical CCSD in a divide-and-conquer manner.
    This starts with an exact rewriting of the per-cell canonical CCSD correlation energy
    \begin{equation}    \label{eq:ecorr_cc}
        E_{\textrm{c}}
            = \frac{1}{N_k} \sum_{ijab} \tau_{iajb} (2V_{iajb} - V_{ibja})
            = \sum_{F}^{n_{\textrm{frag}}} E_{\textrm{c}}^{(F)}
    \end{equation}
    where $\tau_{iajb} = t_{iajb} - t_{ia} t_{jb}$ and the summation is over supercell HF orbitals.
    In the second equality of \cref{eq:ecorr_cc}, we have used \cref{eq:lo_span_occ} to rewrite $E_{\textrm{c}}$ as a summation over fragment contributions,
    \begin{equation}
        E_{\textrm{c}}^{(F)}
            = \sum_{\alpha \in F} \sum_{ii'} U_{i\alpha}^* \mathcal{E}_{ii'} U_{i'\alpha}
    \end{equation}
    where $\mathcal{E}_{ii'}$ depends on the canonical CC amplitudes
    \begin{equation}
        \mathcal{E}_{ii'}
            = \sum_{jab} \tau_{iajb} (2V_{i'ajb} - V_{i'bja})
    \end{equation}
    This alternative way of calculating the CCSD correlation energy (\ref{eq:ecorr_cc}) is exact but provides no cost savings as written because of the need for global CC amplitudes $\tau_{iajb}$.

    LNO-CCSD bypasses the need for global CC amplitudes through a local approximation for the fragment correlation energy, $E_{\textrm{c}}^{(F)}$.
    For each fragment $F$, a local active space $\mathcal{A}_{F}$ is introduced by first transforming the canonical occupied and virtual orbitals separately into a representation optimized for evaluating $E_{\textrm{c}}^{(F)}$.
    Then, only a subset of the transformed orbitals that contribute most significantly to $E_{\textrm{c}}^{(F)}$ are included in $\mathcal{A}_F$, while the rest are kept frozen.
    The exact procedure for the active space construction will be detailed in \cref{subsec:local_active_space}.
    A local Hamiltonian is then defined by constructing the Hamiltonian in the active space $\mathcal{A}_F$,
    \begin{equation}    \label{eq:local_ham}
    \begin{split}
        \hat{H}^{(F)}
            &= \sum_{\sigma} \sum_{pq \in \mathcal{A}_{F}}
                f_{pq}^{(F)} c_{p\sigma}^{\dagger} c_{q\sigma} \\
            &\hspace{1em} +
            \frac{1}{2} \sum_{\sigma\sigma'}\sum_{pqrs \in \mathcal{A}_{\alpha}} V_{pqrs}
            c_{p\sigma}^{\dagger} c_{r\sigma'}^{\dagger} c_{s\sigma'} c_{q\sigma}
    \end{split}
    \end{equation}
    where
    \begin{equation}    \label{eq:local_ham_fock}
        f_{pq}^{(F)}
            = h_{pq} + \sum_{i \notin \mathcal{A}_{F}} (2V_{pqii} - V_{piiq})
    \end{equation}
    Importantly, $\hat{H}^{(F)}$ contains at most two-body interactions and can thus be treated by most existing correlated wavefunction methods.
    In LNO-CCSD, one uses CCSD with the fragment Hamiltonian $\hat{H}^{(F)}$ to obtain $t^{(F)}_{ia}$ and $t^{(F)}_{iajb}$ within $\mathcal{A}_{F}$.
    The fragment amplitudes are then used to obtain a local approximation for $E_{\textrm{c}}^{(F)}$ that can be evaluated completely within $\mathcal{A}_{F}$,
    \begin{equation}    \label{eq:ecorr_cc_LO}
        E_{\textrm{c}}^{(F)}
            \approx \sum_{\alpha \in F} \sum_{ii' \in \mathcal{A}_F}
            U_{i\alpha}^* \mathcal{E}_{ii'}^{(F)} U_{i'\alpha}
    \end{equation}
    where
    \begin{equation}
        \mathcal{E}_{ii'}^{(F)}
            = \sum_{jab \in \mathcal{A}_{F}}
            \tau^{(F)}_{iajb} (2V_{i'ajb} - V_{i'bja})
    \end{equation}
    In LNO-CCSD(T), the additional contribution to $E_{\textrm{c}}$ from the perturbative treatment of triple excitations is approximated in a completely analogous manner
    \begin{equation}    \label{eq:ecorr_t_LO}
        E_{\textrm{c,(T)}}
            \approx \sum_{F}^{n_{\textrm{frag}}} E_{\textrm{c,(T)}}^{(F)}
    \end{equation}
    where
    \begin{equation}
        E_{\textrm{c,(T)}}^{(F)}
            = \sum_{\alpha \in F} \sum_{ii' \in \mathcal{A}_{F}}
            U_{i\alpha}^* \mathcal{E}_{ii'\textrm{,(T)}}^{(F)} U_{i'\alpha}
    \end{equation}
    \begin{equation}
        \mathcal{E}_{ii'\textrm{,(T)}}^{(F)}
            = -\frac{1}{3} \sum_{jkabc \in \mathcal{A}_{F}}
            (4w_{ijk}^{abc} + w_{ijk}^{bca} + w_{ijk}^{cab})
            (v_{i'jk}^{abc} - v_{i'jk}^{cba})
    \end{equation}
    and the intermediates $w_{ijk}^{abc}$ and $v_{ijk}^{abc}$ are defined in section~S1.

    In the limit where $\mathcal{A}_{F}$ includes all orbitals in the HF reference, both the LNO-CCSD and the LNO-CCSD(T) correlation energies [\cref{eq:ecorr_cc_LO,eq:ecorr_t_LO}] converge to the canonical CCSD and CCSD(T) results.
    Therefore, the practicality of the LNO-CC approximation relies on achieving high accuracy with relatively small $\mathcal{A}_{F}$, which depends crucially on the orbitals in $\mathcal{A}_{F}$, as will be discussed in \cref{subsec:local_active_space}.
    A simple composite correction evaluated at the MP2 level can often expedite the convergence, i.e.,
    \begin{equation}    \label{eq:mp2_comp_corr}
        \Delta E_{\textrm{c,MP2}}
            = E_{\textrm{c,MP2}} - \sum_{F}^{n_{\textrm{frag}}} E_{\textrm{c,MP2}}^{(F)}
    \end{equation}
    where $E_{\textrm{c,MP2}}$ is the full-system MP2 correlation energy and $E_{\textrm{c,MP2}}^{(F)}$ is the MP2 fragment correlation energy evaluated with the same $\mathcal{A}_F$ as used in the LNO-CC calculation.

    \subsection{Local active space}
    \label{subsec:local_active_space}

    For each fragment $F$, we partition the HF orbital space into two parts, internal and external.
    The local active space will include all internal orbitals and a compressed subset of the external orbitals.
    The internal and external orbital spaces are based on a singular value decomposition of the LO coefficient matrix
    \begin{subequations}    \label{eq:svd_lo_coeff}
    \begin{align}
        &\mat{U}^{(F)}_{\textrm{occ}}
            = \begin{bmatrix}
                \mat{W}^{(F)}_{\textrm{occ}} & \bar{\mat{W}}^{(F)}_{\textrm{occ}}
            \end{bmatrix}
            \begin{bmatrix}
                \bm{\Sigma}^{(F)}_{\textrm{occ}} & \\
                & \bm{0}
            \end{bmatrix}
            \begin{bmatrix}
                \mat{V}^{(F)}_{\textrm{occ}} & \bar{\mat{V}}^{(F)}_{\textrm{occ}}
            \end{bmatrix}^{\dagger} \\
        &\mat{U}^{(F)}_{\textrm{vir}}
            = \begin{bmatrix}
                \mat{W}^{(F)}_{\textrm{vir}} & \bar{\mat{W}}^{(F)}_{\textrm{vir}}
            \end{bmatrix}
            \begin{bmatrix}
                \bm{\Sigma}^{(F)}_{\textrm{vir}} & \\
                & \bm{0}
            \end{bmatrix}
            \begin{bmatrix}
                \mat{V}^{(F)}_{\textrm{vir}} & \bar{\mat{V}}^{(F)}_{\textrm{vir}}
            \end{bmatrix}^{\dagger}
    \end{align}
    \end{subequations}
    where $\mat{U}^{(F)}_{\textrm{occ}}$ is an $N_{\textrm{occ}} \times n_{\textrm{LO}}^{(F)}$ submatrix of $\mat{U}$ where the rows correspond to the $N_{\textrm{occ}}$ occupied orbitals and the columns correspond to the $n_{\textrm{LO}}^{(F)}$ fragment LOs,
    and $\mat{U}^{(F)}_{\textrm{vir}}$ is defined similarly for the virtual orbitals.
    The internal space consists of the occupied and virtual orbitals defined by the left singular vectors with non-zero singular values
    \begin{subequations}    \label{eq:internal_orbitals}
    \begin{align}
        &\psi_{i_{F}}
            = \sum_{j} \psi_{j} W_{ji_F}^{(F)},
        \quad{}i_F = 1, \cdots, n_{\textrm{int-occ}}^{(F)}  \\
        &\psi_{a_{F}}
            = \sum_{b} \psi_{b} W_{ba_F}^{(F)},
        \quad{}a_F = 1, \cdots, n_{\textrm{int-vir}}^{(F)}
    \end{align}
    \end{subequations}
    where both $n_{\textrm{int-occ}}^{(F)}$ and $n_{\textrm{int-vir}}^{(F)}$ are no greater than $n_{\textrm{LO}}^{(F)}$.
    These orbitals are entangled with the fragment LOs and thus included in $\mathcal{A}_F$.
    The remaining orbitals, defined by the left singular vectors with zero singular values, make up the external space.
    Note that for PM orbitals that only span the occupied space, the internal virtual space vanishes and all virtual orbitals are external.

    The external orbitals, although disentangled from the fragment LOs, contribute significantly to the dynamic correlation energy in $E^{(F)}_{\textrm{c}}$.
    Therefore, they must be carefully compressed to balance accuracy and computational efficiency.
    In LNO-CC, we use the LNOs calculated at MP2 level to compress the external space.
    First, we calculate the occupied and virtual blocks of the semi-canonical MP2 density matrix, restricting the summation of one occupied index to be within internal space,
    \begin{subequations}    \label{eq:mp2_1rdm_lo}
    \begin{align}
        &D_{ij}^{(F)}
            = 2 \sum_{k_{F}}^{n_{\textrm{int-occ}}^{(F)}} \sum_{ab}
            t^{(1)*}_{iak_{F}b} (
                2 t^{(1)}_{jak_{F}b}
                - t^{(1)}_{jbk_{F}a})    \\
        &D_{ab}^{(F)}
            = \sum_{i_{F}}^{n_{\textrm{int-occ}}^{(F)}} \sum_{jc}
            2 \left(
                t^{(1)}_{i_{F}ajc} t^{(1)*}_{i_{F}bjc}
              + t^{(1)}_{i_{F}cja} t^{(1)*}_{i_{F}cjb}
            \right) \nonumber \\
        &\phantom{D_{ab,F}=\sum_{i_{F}}^{n_{\textrm{LO},F}} \sum_{jc}}
            - \left(
                t^{(1)}_{i_{F}cja} t^{(1)*}_{i_{F}bjc}
              + t^{(1)}_{i_{F}ajc} t^{(1)*}_{i_{F}cjb}
            \right)
    \end{align}
    \end{subequations}
    where
    \begin{equation}    \label{eq:mp2_amp}
        t^{(1)}_{i_{F}ajb}
            = \frac{V_{i_{F}ajb}^*}
            {f_{i_{F}i_{F}} + \varepsilon_{j} - \varepsilon_a - \varepsilon_b}
    \end{equation}
    are semi-canonical MP1 amplitudes~[REF].
    The ERIs in \cref{eq:mp2_amp} can be efficiently approximated using density fitting (DF)~\cite{Ye21JCPa,Ye21JCPb}
    \begin{equation}    \label{eq:eri_mp2}
        V_{i_{F}ajb}
            = \sum_{\bm{q}}^{N_k} \sum_{P}^{n_{\textrm{aux}}}
            \sum_{i} W^*_{ii_F} L_{i a}^{P}(\bm{q}) L_{jb}^{P}(-\bm{q})
    \end{equation}
    where $P$ labels a set of $n_{\textrm{aux}}$ auxiliary basis functions, and
    \begin{equation}    \label{eq:eri_mp2_df3c}
        L_{i a}^{P}(\bm{q})
            = \sum_{\bm{k}_1\bm{k}_2}^{N_k} \delta_{\bm{k}_{12},\bm{q}}
            \sum_{\mu\nu} L_{\mu\bm{k}_1\nu\bm{k}_2}^{P\bm{k}_{12}}
            C_{\mu\bm{k}_1 i}^* C_{\nu\bm{k}_2 a}
    \end{equation}
    where $L_{\mu\bm{k}_1\nu\bm{k}_2}^{P\bm{k}_{12}}$ are the DF factors in the atomic orbital (AO) basis, $\bm{k}_{12} = \bm{k}_2-\bm{k}_1$, and $C_{\mu\bm{k} p}$ are expansion coefficients of the supercell HF orbitals in the $k$-point AO basis.
    Diagonalizing the MP2 density matrix (\ref{eq:mp2_1rdm_lo}) within the external space
    \begin{subequations}
    \begin{align}
        &\bar{\mat{D}}^{(F)}_{\textrm{occ}}
            = \bar{\mat{W}}^{(F)\dagger}_{\textrm{occ}}
            \mat{D}^{(F)}_{\textrm{occ}}
            \bar{\mat{W}}^{(F)}_{\textrm{occ}}
            = \mat{X}^{(F)}_{\textrm{occ}}
            \bm{\Lambda}_{\textrm{occ}}
            \mat{X}^{(F)\dagger}_{\textrm{occ}} \\
        &\bar{\mat{D}}^{(F)}_{\textrm{vir}}
            = \bar{\mat{W}}^{(F)\dagger}_{\textrm{vir}}
            \mat{D}^{(F)}_{\textrm{vir}}
            \bar{\mat{W}}^{(F)}_{\textrm{vir}}
            = \mat{X}^{(F)}_{\textrm{vir}}
            \bm{\Lambda}_{\textrm{vir}}
            \mat{X}^{(F)\dagger}_{\textrm{vir}}
    \end{align}
    \end{subequations}
    leads to the MP2 LNOs
    \begin{subequations}    \label{eq:lno}
    \begin{align}
        &\xi_{i_F}
            = \sum_{j} \psi_{j} \left[
                \bar{\mat{W}}_{\textrm{occ}}^{(F)} \mat{X}_{\textrm{occ}}^{(F)}
            \right]_{ji_F}
        \\
        &\xi_{a_F}
            = \sum_{b} \psi_{b} \left[
                \bar{\mat{W}}_{\textrm{vir}}^{(F)} \mat{X}_{\textrm{vir}}^{(F)}
            \right]_{ba_F}
    \end{align}
    \end{subequations}
    where the eigenvalues $\lambda_{p_F}^{(F)}$ quantify the importance of the corresponding LNOs to $E^{(F)}_{\textrm{c}}$.
    We include in $\mathcal{A}_F$ the occupied and virtual LNOs with significant eigenvalues
    \begin{equation}    \label{eq:lno_thresh}
        |\lambda_{i_F}^{(F)}|
            \geq \eta_{\textrm{occ}},
        \quad{}
        |\lambda_{a_F}^{(F)}|
            \geq \eta_{\textrm{vir}}
    \end{equation}
    for some user-defined thresholds $\eta_{\textrm{occ}}$ and $\eta_{\textrm{vir}}$.

\begin{figure}[!th]
        \centering
        \includegraphics[width=7cm]{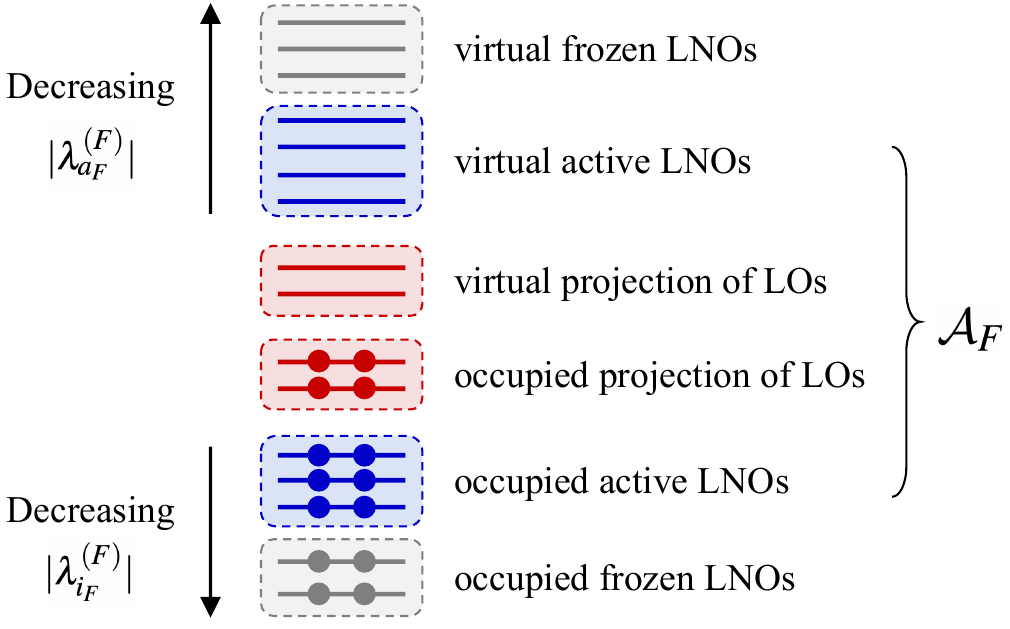}
        \caption{Schematic illustration of a local active space generated for an IAO-based fragment.
        The canonical HF occupied and virtual orbitals are rotated separately into six subsets: the occupied and virtual projection of the LOs (red), the occupied and virtual active LNOs (blue), and the occupied and virtual frozen LNOs (gray).
        The local active space $\mathcal{A}_F$ comprises all non-frozen orbitals.
        For fragment consisting of PM orbitals, there is no virtual projection of LOs.}
        \label{fig:las}
    \end{figure}

    To summarize, for each fragment $F$, the HF orbitals are transformed by fragment-specific unitary rotations into three subspaces: the internal space comprising orbitals entangled with the fragment LOs (\ref{eq:internal_orbitals}), the active external space comprising LNOs with significant eigenvalues (\ref{eq:lno_thresh}), and the frozen external space comprising the remaining LNOs with negligible eigenvalues.
    The local active space $\mathcal{A}_F$ includes all orbitals from the internal and the active external subspaces, containing $n_{\textrm{occ}}^{(F)} + n_{\textrm{vir}}^{(F)} = n_{\textrm{MO}}^{(F)}$ orbitals in total.
    This is also illustrated pictorially in \cref{fig:las}.
    As mentioned in \cref{subsec:periodic_lno_cc}, the HF reference remains unchanged after the unitary rotations because they are separately applied to the occupied and the virtual spaces.

    \subsection{Computational cost}
    \label{subsec:computational_cost}

    Let $n$ and $N$ denote quantities that scale with the size of the unit cell and the supercell, respectively.
    Let $m$ denote the size of a typical local active space.
    The CPU cost of a periodic LNO-CC calculation is dominated by three parts:
    \begin{enumerate}
        \item Calculating the MP2 density matrix (\ref{eq:mp2_1rdm_lo}) needed for generating LNOs, whose cost scales as $\mathcal{O}(N_k^4 n_\mathrm{MO}^4)$ per fragment.
        \item Transforming ERIs for the local Hamiltonian (\ref{eq:local_ham}), which can be performed efficiently using DF similar to \cref{eq:eri_mp2,eq:eri_mp2_df3c}
        \begin{subequations}    \label{eq:eri_local_ham}
        \begin{align}
            &V_{pqrs}
                = \sum_{\bm{q}}^{N_k} \sum_{P}^{n_{\textrm{aux}}}
                L_{pq}^{P}(\bm{q}) L_{rs}^{P}(-\bm{q})  \label{eq:eri_local_ham_4c}
            \\
            &L_{pq}^{P}(\bm{q})
                = \sum_{\bm{k}_1\bm{k}_2}^{N_k} \delta_{\bm{k}_{12},\bm{q}}
                \sum_{\mu\nu} L_{\mu\bm{k}_1\nu\bm{k}_2}^{P\bm{k}_{12}}
                C_{\mu\bm{k}_1 p}^* C_{\nu\bm{k}_2 q}   \label{eq:eri_local_ham_3c}
        \end{align}
        \end{subequations}
        for $p,q,r,s \in \mathcal{A}_F$.
        The cost of evaluating \cref{eq:eri_local_ham_4c,eq:eri_local_ham_3c} per fragment scales as $\mathcal{O}[N_k  n_{\textrm{aux}}  (n_{\textrm{MO}}^{(F)})^4]$ and $\mathcal{O}\{N_k^2  n_{\textrm{aux}} [n_\mathrm{AO}^2 n_{\textrm{MO}}^{(F)}  + n_{\textrm{AO}} (n_{\textrm{MO}}^{(F)})^2]\}$, respectively.
        \item Solving the CC amplitude equations in $\mathcal{A}_F$ with the local Hamiltonian (\ref{eq:local_ham}), whose cost per fragment is $\mathcal{O}[(n_{\textrm{MO}}^{(F)})^6]$ and $\mathcal{O}[(n_{\textrm{MO}}^{(F)})^7]$ for LNO-CCSD and LNO-CCSD(T), respectively.
        \item Calculating the MP2 composite correction (\ref{eq:mp2_comp_corr}), which requires a full MP2 calculation that scales as $\mathcal{O}(N_k^3 n_{\textrm{MO}}^4 n_{\textrm{aux}})$.
    \end{enumerate}
    In addition to the CPU cost, our current implementation also requires storing the DF tensors $L_{\mu\bm{k}_1\nu\bm{k}_2}^{P\bm{k}_{12}}$ and $L_{ia}^{P}(\bm{q})$ ($i,a$ being supercell HF orbitals) on disk,
    which requires storage of $\mathcal{O}(N_k^2 n_{\textrm{AO}}^2 n_{\textrm{aux}})$ and $\mathcal{O}(N_k^3 n_{\textrm{occ}} n_{\textrm{vir}} n_{\textrm{aux}})$, respectively.

    \subsection{Relation to other methods}
    \label{subsec:relation_to_others}

    As a local active space method, periodic LNO-CC is related to many existing quantum embedding methods.
    The singular value decomposition of the overlap matrix between the fragment LOs and the occupied/virtual HF orbitals (\ref{eq:svd_lo_coeff}) is equivalent to the Schmidt decomposition of the HF reference wavefunction used in density matrix embedding theory~\cite{Knizia13JCTC,Wouters16JCTC} (DMET) and related methods~\cite{Bulik14JCP,Ye19JCTC}.
    In DMET, the internal space is augmented with projected atomic orbitals~\cite{Boughton93JCC} (PAOs) selected from the external space to make the local active space~\cite{Cui20JCTC}.
    Both the Schmidt decomposition and the PAO generation can be performed at mean-field cost, which is cheaper than the use of MP2 LNOs.
    However, PAOs are known to be less effective at capturing dynamic correlation compared to the MP2 LNOs~\cite{Yang11JCP}.
    As a result, the convergence of the DMET correlation energy with fragment size can be slow in large basis sets~\cite{Ye19JPCL}.

    A recent extension of DMET~\cite{Nusspickel22PRX,Nusspickel23JCTC} replaces PAOs with LNOs for the external space, which leads to significantly faster convergence of the correlation energy compared to the original DMET.
    The LNOs used in Ref.~\citenum{Nusspickel22PRX} are generated using slightly different MP2 density matrices compared to those used in this work [\cref{eq:mp2_1rdm_lo}],
    \begin{subequations}    \label{eq:mp2_1rdm_lo_2h2p}
    \begin{align}
        &D_{ij}^{(F)}
            = 2 \sum_{a_F b_F}^{n_{\textrm{int-vir}}^{(F)}} \sum_{k}
            t^{(1)*}_{iakb} (
                2 t^{(1)}_{ja_{F}kb_{F}}
                - t^{(1)}_{jb_{F}ka_{F}})   \label{eq:mp2_1rdm_lo_2h2p_occ}
        \\
        &D_{ab}^{(F)}
            = 2 \sum_{i_{F} j_{F}}^{n_{\textrm{int-occ}}^{(F)}} \sum_{c}
            t^{(1)*}_{i_{F}aj_{F}c} (
                2 t^{(1)}_{i_{F}bj_{F}c}
                - t^{(1)}_{i_{F}cj_{F}b})   \label{eq:mp2_1rdm_lo_2h2p_vir}
    \end{align}
    \end{subequations}
    where the canonical and internal orbitals being summed to calculate the density matrix are flipped; these LNOs were called cluster-specific bath natural orbitals (CBNOs) in Ref.~\citenum{Nusspickel23JCTC}.
    The use of virtual internal orbitals in \cref{eq:mp2_1rdm_lo_2h2p_occ} means that only LOs overlapping with the virtual space, e.g.,~IAOs, can be used in the construction of occupied CBNOs.
    Due to using more internal orbitals than external orbitals in calculating the MP2 density matrices, the computational cost of generating the CBNOs (\ref{eq:mp2_1rdm_lo_2h2p}) is $\mathcal{O}(N_k^3 n_\mathrm{MO}^3)$ per fragment, which is lower than the $\mathcal{O}(N_k^4 n_\mathrm{MO}^4)$ cost scaling of generating the LNOs (\ref{eq:mp2_1rdm_lo}) used in this work.
    We will compare the accuracy of CBNO-CC and LNO-CC in \cref{sec:results_and_discussion}.



    \section{Computational details}
    \label{sec:comput_details}

    All calculations presented below are performed using a development version of PySCF 2.5~\cite{Sun20JCP}, which uses Libcint~\cite{Sun19JCC} for Gaussian integral evaluation.
    Correlation-consistent Gaussian basis sets~\cite{Ye22JCTC} optimized for the Goedecker-Teter-Hutter (GTH) pseudopotential~\cite{Goedecker96PRB,Hartwigsen98PRB,HutterPP} are used as the AO basis.
    ERIs are approximated using range-separated density fitting~\cite{Ye21JCPa,Ye21JCPb}, with optimized fitting basis sets.
    The integrable divergence of the HF exchange integral is treated with a Madelung constant correction~\cite{Paier06JCP,Broqvist09PRB,Sundararaman13PRB}; although this correction does not affect canonical CC results, it does affect MP2 results and thus affects LNO-CC through the LNOs and the composite correction.
    Two bulk crystals, diamond and BCC lithium, will be used to test the performance of LNO-CC.
    The convergence of LNO-CC to the canonical limit with respect to the number of LNOs in each fragment is monitored by scanning $\eta_{\textrm{vir}}$ while fixing the ratio $\gamma = \eta_{\textrm{occ}}/\eta_{\textrm{vir}}$.
    For diamond, we follow the recommendations of molecular LNO-CC~\cite{Rolik13JCP} and use $\gamma = 10$.
    For lithium, we use $\gamma = 1$ based on our own preliminary testing.
    As explained in \cref{subsec:relation_to_others}, our LNO-CC code can straightforwardly perform CBNO-CC calculations, whose results will also be included for comparison.
    Preliminary testing suggests that the CBNO-CC results show little dependence on the choice of $\gamma$ within the range of $0.1$ to $10$.
    We thus use $\gamma = 1$ for CBNO-CC to be consistent with the original paper~\cite{Nusspickel22PRX}.
    All calculations were performed on a single node with 24 CPU cores (AMD EPYC 7302, 2.99 GHz) and 8 GB of memory per core.

    \section{Results and discussion}
    \label{sec:results_and_discussion}

    \subsection{Diamond}
    \label{subsec:diamond}

    \begin{figure}[!th]
        \centering
        \includegraphics[width=8cm]{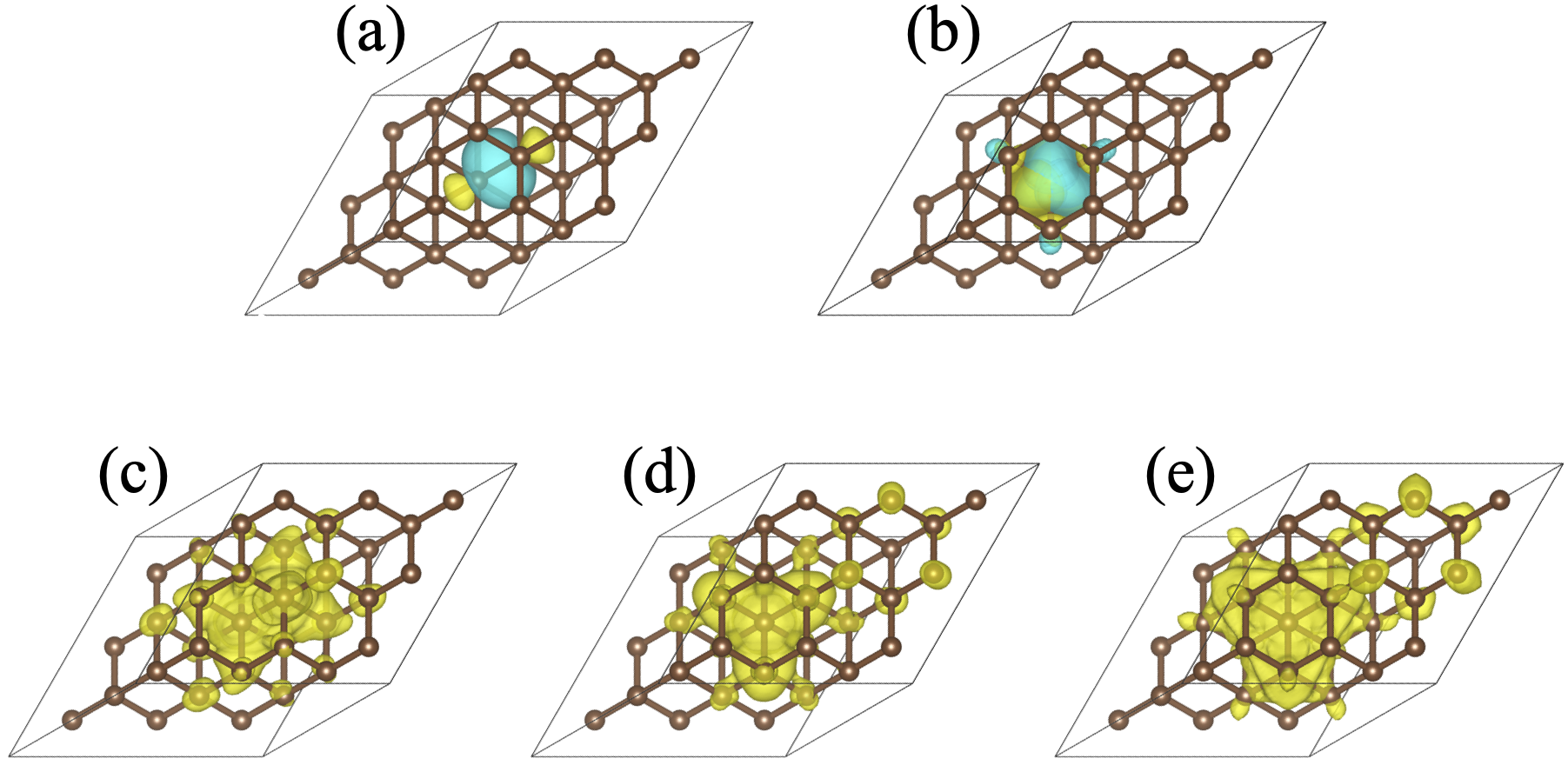}
        \caption{LOs and the density of the associated active-space orbitals for one representative fragment in a $3\times3\times3$ supercell of diamond.
        (a) A PM orbital localized on a \ce{C-C} bond.
        (b) Overlay of four IAOs ($2s,~2p_x,~2p_y$ and $2p_z$) localized on a carbon atom.
        (c--e) The charge density of $300$ virtual active-space orbitals in (c) LNO-CC/PM, (d) LNO-CC/IAO, and (e) CBNO-CC/IAO, visualized using the same isosurface value ($0.7$~a.u.).
        }
        \label{fig:diamond_orbs}
    \end{figure}

    Our first test system is diamond crystal, a covalent insulator with a large band gap and hence relatively local electronic structure.
    In \cref{fig:diamond_orbs}a and b, we show a representative PM orbital localized on a \ce{C-C} single bond and the IAOs localized on a single carbon atom within a $3\times3\times3$ supercell.
    The high symmetry of the lattice indicates that all fragments generated based on either type of LOs are equivalent and contribute equally to the correlation energy.
    As a result, for both choices of LOs, only a single fragment calculation is needed, containing one (PM) or four (IAO) LOs (when using a pseudopotential for carbon).

    \begin{figure}[!th]
        \centering
        \includegraphics[width=8cm]{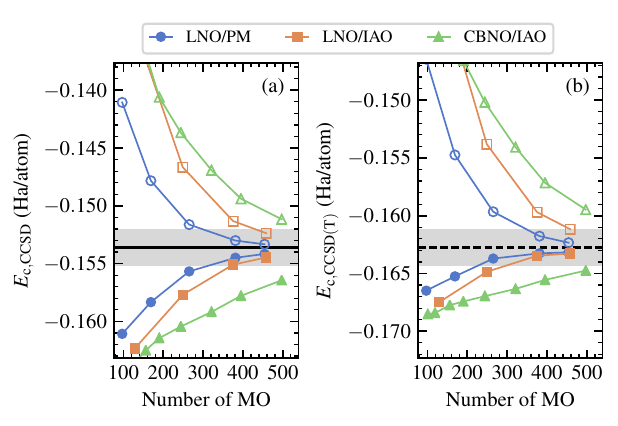}
        \caption{Comparing the convergence of CCSD (left) and CCSD(T) (right) correlation energy with the local active space size evaluated using LNO-CC and CBNO-CC with different LO types for diamond at its experimental geometry ($a = 3.567~\textrm{\AA}$) using a TZ basis set, a GTH pseudopotential, and a $3\times3\times3$ $k$-point mesh.
        Hollow and filled markers correspond to results without and with the MP2 composite correction (\ref{eq:mp2_comp_corr}), respectively.
        The canonical $k$-point CCSD energy is shown as a solid black horizontal line in panel (a).
        Our best estimate of the converged CCSD(T) energy [obtained by averaging the uncorrected and corrected LNO-CCSD(T)/PM results with the largest active space size of approximately $400$ orbitals], is shown as a dashed black horizontal line in panel (b).
        The gray shaded area indicates a range of $\pm~1$~kcal/mol from the chosen reference.
        }
        \label{fig:diamond_eq_conv}
    \end{figure}

    We first investigate how different choices of LOs influence the accuracy of the LNO-CC correlation energy.
    In \cref{fig:diamond_eq_conv}a and b, we show the LNO-CCSD and LNO-CCSD(T) correlation energies for diamond at its experimental geometry ($a = 3.567$~\AA{}) as a function of the number of active-space orbitals, evaluated using the TZ basis set with the GTH pseudopotential and a $3\times3\times3$ $k$-point mesh to sample the Brillouin zone.
    Results are shown for both PM orbitals and IAOs, with and without the MP2 composite correction (\ref{eq:mp2_comp_corr}).
    There are $58$ HF orbitals per $k$-point in the chosen basis, resulting in a total of $1566$ orbitals in the $3\times 3\times 3$ supercell.
    Canonical CCSD is feasible by exploiting $k$-point symmetry for this system size [giving $\mathcal{O}(N_k^4 n_\mathrm{MO}^6)$ scaling], and the result is represented by the solid black horizontal line in \cref{fig:diamond_eq_conv}a.
    A corresponding canonical CCSD(T) calculation is infeasible even for this system size, and our best estimate---obtained by averaging the corrected and uncorrected LNO-CCSD(T)/PM results using about $460$ orbitals---is represented by the dashed black horizontal line in \cref{fig:diamond_eq_conv}b.
    In both cases, the gray shaded area depicts a range of $\pm 1$~kcal/mol from the chosen reference.

    When the MP2 correction is not applied, LNO-CC/PM requires approximately $250$ orbitals to achieve an accuracy of $1$~kcal/mol in the CCSD correlation energy and about $300$ orbitals for the CCSD(T) correlation energy.
    In contrast, when IAOs are used instead of PM orbitals, around $400$ orbitals are needed for the same level of accuracy for CCSD, and $450$ orbitals for CCSD(T).
    In all cases, LNO-CC underestimates the magnitude of the correlation energy and the convergence to the reference is from above.
    Including the MP2 correction results in an overestimation of the correlation energy, causing convergence from below, but otherwise reduces the absolute error in most cases.
    The improvement is more significant for LNO-CCSD(T) than LNO-CCSD, and for IAOs than PM orbitals, which leads to a reduced difference between IAOs and PM orbitals especially for LNO-CCSD(T).

    In \cref{fig:diamond_eq_conv}, we also present results from CBNO-CC/IAO for comparison.
    These results exhibit slower convergence with respect to active space size compared to LNO-CC, regardless of the choice of LOs, and for both CCSD and CCSD(T).
    As explained in \cref{subsec:relation_to_others}, the primary difference between LNO-CC/IAO and CBNO-CC/IAO stems from the type of MP2 LNOs used to compress the external space.
    In particular, the LNOs used in LNO-CC/IAO are computationally more expensive to evaluate.
    The results in \cref{fig:diamond_eq_conv} thus suggest that the higher cost of LNO construction in LNO-CC is outweighed by the smaller number of active space orbitals required for a certain accuracy.
    A separate calculation with occupied LNOs from LNO-CC/IAO but different virtual LNOs from either LNO-CC/IAO or CBNO-CC/IAO reveals that the virtual LNOs are responsible for the difference (fig.~S3).
    To gain some insight, we show in \cref{fig:diamond_orbs}c--e the charge density of $300$ virtual orbitals in the local active space of LNO-CC/PM, LNO-CC/IAO and CBNO-CC/IAO, calculated by
    \begin{equation}    \label{eq:lno_charge_density}
        \rho(\bm{r})
            = 2 \sum_{a=1}^{300} |\xi_{a}(\bm{r})|^2
    \end{equation}
    While the virtual LNOs correctly localize around the underlying LOs in all three cases, the virtual LNO density from CBNO-CC/IAO is clearly more diffuse compared to the corresponding ones from LNO-CC/IAO.
    This comparison highlights the subtlety of choosing an appropriate flavor of LNOs.

    \begin{figure}[!th]
        \centering
        \includegraphics[width=8cm]{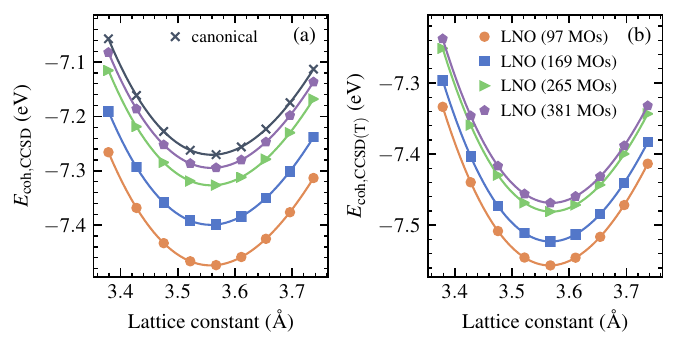}
        \caption{Convergence of CCSD (left) and CCSD(T) (right) equation of state with the local active space size evaluated using LNO-CC with PM LOs for diamond in a TZ basis set with the GTH pseudopotential.
        The curves are from fitting the data using Birch-Murnaghan equation.
        The LNO-CC results are obtained using a $3\times3\times3$ $k$-point mesh and corrected for finite-size effects at the MP2 level.
        Results from canonical CCSD are shown in black in panel (a).
        }
        \label{fig:diamond_eos_conv}
    \end{figure}

    We then study the equation of state (EOS) of diamond near the equilibrium geometry.
    In \cref{fig:diamond_eos_conv}, we present the LNO-CCSD and LNO-CCSD(T) cohesive energies, $E_{\textrm{coh}}$, obtained by subtracting the counterpoise-corrected single-atom CCSD and CCSD(T) energies from the LNO-CCSD and LNO-CCSD(T) lattice energies, as a function of the lattice constant.
    These calculations are performed using PM orbitals and four different active space sizes, all with the MP2 composite correction.
    The remaining finite-size errors due to using a $3\times3\times3$ $k$-point mesh and basis set incompleteness errors from using a TZ basis set are corrected at the MP2 level, as described in section~S2.4.
    From \cref{fig:diamond_eos_conv}, we observe that smooth energy curves with qualitatively correct curvature are obtained for all active space sizes.
    For CCSD, where the canonical reference is available, the convergence of the LNO-CCSD energy curves to the reference curve closely follows the pattern observed in \cref{fig:diamond_eq_conv}a at the equilibrium geometry.
    Remarkably, an accuracy of $0.06$ and $0.02$~eV in the equilibrium cohesive energy is achieved using $265$ and $381$ active space orbitals, respectively, which are a small fraction of the total orbital count, $1566$.
    The convergence of the LNO-CCSD(T) energy curves is even faster compared to LNO-CCSD.
    Similar results obtained using IAOs are shown in fig.~S3, which exhibit a convergence pattern similar to that of \cref{fig:diamond_eos_conv} but require, on average, $100$ more active orbitals to achieve the same level of accuracy, which is consistent with the pattern observed in \cref{fig:diamond_eq_conv}.

    \begin{figure}[!th]
        \centering
        \includegraphics[width=8cm]{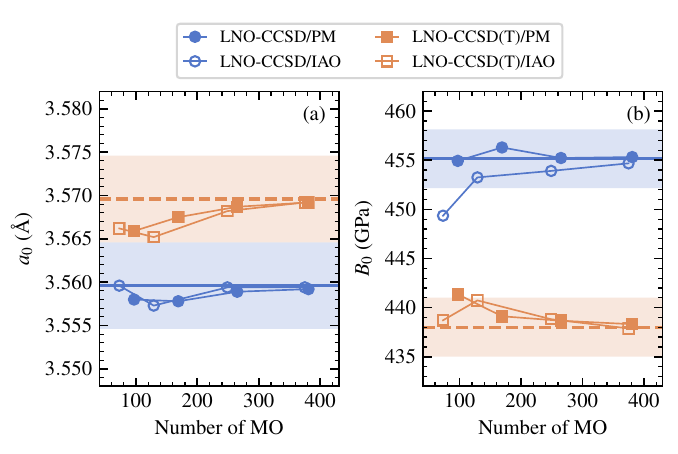}
        \caption{Convergence of the lattice constant ($a_0$, left) and bulk modulus ($B_0$, right) predicted by LNO-CCSD and LNO-CCSD(T) with different LOs for diamond using a TZ basis set with the GTH pseudopotential.
        The LNO-CC energies are first evaluated using the TZ basis and a $3\times3\times3$ $k$-point mesh, and basis set and finite-size errors are corrected at the MP2 level.
        For CCSD, reference results from canonical CCSD are shown as blue solid lines.
        For CCSD(T), reference results from LNO-CCSD(T)/PM with approximately $500$ active orbitals are shown as orange dashed lines.
        The shaded area highlights errors less than $\pm 0.005$~\AA{} in $a_0$ and $\pm 3$~GPa in $B_0$.
        }
        \label{fig:diamond_a0b0_conv}
    \end{figure}

    Fitting the energy curves with the Birch-Murnaghan equation of state allows us to extract the equilibrium lattice constant $a_0$ and bulk modulus $B_0$ predicted by LNO-CCSD and LNO-CCSD(T).
    In \cref{fig:diamond_a0b0_conv}, we show the convergence of the two structural parameters with active space size.
    We see that a plateau is quickly achieved for both parameters by both LNO-CCSD and LNO-CCSD(T).
    For CCSD, relatively small errors of $\pm 0.005$~\AA{} in $a_0$ and $\pm 3$~GPa in $B_0$ from the canonical reference (highlighted by the shaded area) are achieved by LNO-CCSD with both PM orbitals and IAOs using only $100$ active space orbitals.
    The same convergence pattern is observed for LNO-CCSD(T), resulting in reliable predictions of the two structural parameters at the CCSD(T) level even when a canonical reference is infeasible.

    \begin{table}[!tbh]
        \centering
        \caption{Thermochemical properties of diamond predicted by different levels of wavefunction theories compared to experiments.}
        \label{tab:diamond_prop}
        \begin{tabular}{lcccl}
            \hline\hline
            Method & $E_{\textrm{coh}}/\textrm{eV}$ & $a_0/\textrm{\AA}$ & $B_0/\textrm{GPa}$ \\
            \hline
            HF/GTH-HF & $5.17$ & $3.557$ & $490$ & \multirow{4}{*}{this work} \\
            MP2/GTH-HF & $7.87$ & $3.554$ & $449$ & \\
            CCSD/GTH-HF & $7.29$ & $3.560$ & $455$ & \\
            CCSD(T)/GTH-HF & $7.47$ & $3.570$ & $438$ & \\
            \hline
            HF/All-e & $5.31$ & $3.550$ & $498$ & \multirow{4}{*}{this work} \\
            MP2/All-e & $8.04$ & $3.547$ & $458$ & \\
            CCSD/All-e & $7.47$ & $3.551$ & $466$ & \\
            CCSD(T)/All-e & $7.64$ & $3.561$ & $448$ & \\
            \hline
            CBNO-CCSD/All-e & & $3.557$ & $467$ & ref~\citenum{Nusspickel22PRX} \\
            \hline
            Experiment & $7.55$ & $3.553$ & $453$ & ref~\citenum{Schimka11JCP} \\
            \hline
        \end{tabular}
    \end{table}

    In \cref{tab:diamond_prop}, we present our final predictions for the equilibrium $E_{\textrm{coh}}$, $a_0$, and $B_0$ of diamond at both the CCSD and CCSD(T) levels, derived from converged LNO-CC calculations using the GTH pseudopotential.
    Overall, the tendency of MP2 to overbind is corrected by CCSD and CCSD(T), resulting in $E_{\textrm{coh}}$ predictions that align more closely with experimental values.
    However, $B_0$ predicted by CCSD(T) is less accurate than that of MP2 and CCSD, which could be due to the pseudopotential approximation.
    To assess pseudopotential errors, we performed additional all-electron LNO-CC calculations using Dunning's cc-pVTZ basis set~\cite{Dunning89JCP} with core electrons frozen, but otherwise following the protocol described above.
    The convergence of these all-electron LNO-CC calculations closely follows the pseudopotential-based calculations and is summarized in section~S2.4.
    The final results of the thermochemical properties of diamond from all-electron HF, MP2, and LNO-CC calculations are also shown in \cref{tab:diamond_prop}.
    We observe that using the all-electron potential results in a consistent shift of all three properties regardless of the level of theory.
    The shifts are approximately $+0.15$~eV for $E_{\textrm{coh}}$, $-0.01$~\AA{} for $a_0$, and $+10$~GPa for $B_0$.
    These shifts do not alter the trends discussed earlier but significantly improve the agreement of the CCSD(T) predicted $B_0$ with experimental data.
    The similarity in the shifts due to the use of pseudopotentials across different levels of theory indicates that the primary error source from the pseudopotential is at the HF level.
    \Cref{tab:diamond_prop} also includes the all-electron CCSD $a_0$ and $B_0$ reported in ref~\citenum{Nusspickel22PRX} using CBNO-CC with a large active space of more than $400$ orbitals, and the results are in nearly perfect agreement with our all-electron CCSD numbers.

    \begin{figure}
        \centering
        \includegraphics[width=5cm]{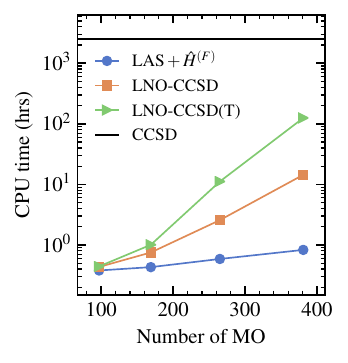}
        \caption{Computational cost of LNO-CCSD and LNO-CCSD(T) as a function of active space size for diamond at its experimental geometry using a TZ basis set and a $3\times3\times3$ $k$-point mesh.
        The component cost of constructing the local active space and local Hamiltonian is shown separately.
        The cost of canonical $k$-point CCSD is indicated by the black horizontal line.
        }
        \label{fig:diamond_time}
    \end{figure}

    Finally, we present in \cref{fig:diamond_time} the computational cost of LNO-CCSD and LNO-CCSD(T) as a function of active space size for diamond at its experimental geometry using a TZ basis set and a $3\times3\times3$ $k$-point mesh.
    For the largest active space of $381$ orbitals where quantitative accuracy is achieved in all tested properties, LNO-CCSD exhibits a 200-fold speedup compared to canonical $k$-point CCSD.
    The speedup increases to about $1000$-fold when using a slightly smaller active space of $265$ orbitals, with only minor loss of accuracy as discussed above.
    Additionally, even LNO-CCSD(T) is faster than canonical CCSD by a factor of about 200 using the $265$-orbital active space (recall that canonical CCSD(T) is infeasible for this problem).
    In \cref{fig:diamond_time}, we also show the combined cost of constructing the local active space and the local Hamiltonian.
    We observe that for active spaces containing more than $200$ orbitals, the cost of LNO-CC is dominated by the fragment CC calculations.
    For larger systems, however, the cost of constructing $\mathcal{A}_{F}$ and $\hat{H}^{(F)}$ will eventually become dominant due to their unfavorable scaling with the supercell size $N_k$.
    Future work will explore the use of local density fitting~\cite{Nagy16JCTC,Nagy19JCTC} to mitigate this cost scaling.

    \subsection{Lithium}
    \label{subsec:lithium}

    Our second test system is BCC lithium crystal.
    In contrast to the local electronic structure of diamond, lithium is a metal characterized by a vanishing band gap in the TDL.
    This leads to a divergent correlation energy from any finite-order perturbation theories, including MP2 and the perturbative triples in CCSD(T)~\cite{Shepherd13PRL,Masios23PRL,Neufeld23PRL}.
    CCSD, however, remains a valid theory for metals even in the TDL.
    The performance of canonical CCSD on the ground-state thermochemical properties of BCC lithium has been benchmarked recently in ref~\citenum{Neufeld22JPCL,Masios23PRL}, showing reasonable agreement with available experimental data.
    Here, we investigate whether LNO-CCSD can effectively approximate canonical CCSD even for this metallic system.

    \begin{figure}[!th]
        \centering
        \includegraphics[width=7cm]{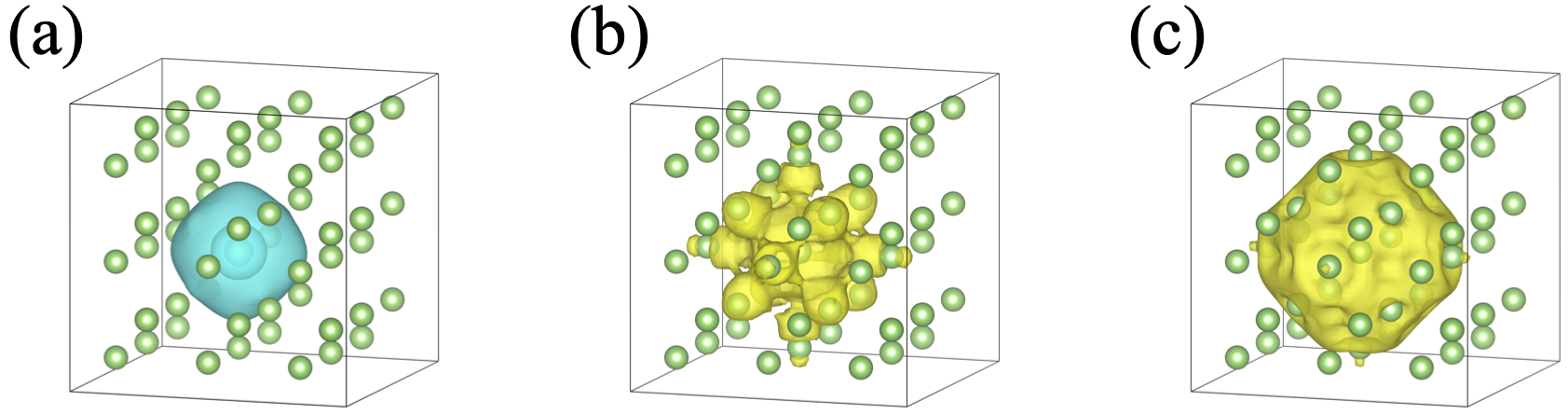}
        \caption{(a) Overlay of two IAOs ($1s$ and $2s$) localized on a lithium atom.
        (b,c) The charge density of $300$ virtual active-space orbitals in (b) LNO-CC/IAO and (c) CBNO-CC/IAO, visualized using the same isosurface value ($0.15$~a.u.).
        }
        \label{fig:lithium_orbs}
    \end{figure}

\begin{figure}[!th]
        \centering
        \includegraphics[width=8.5cm]{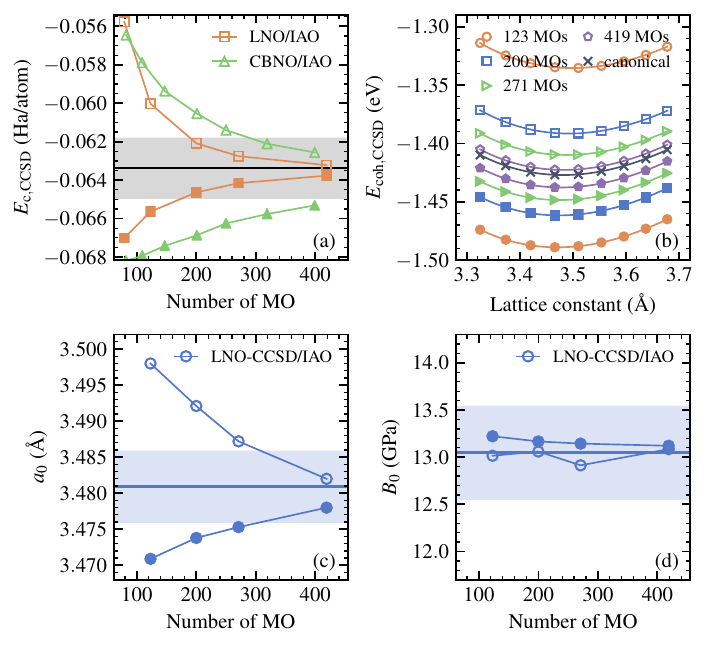}
        \caption{(a) Same plot as \cref{fig:diamond_eq_conv}a for BCC lithium using a TZ basis set with the GTH pseudopotential and a $3\times3\times3$ $k$-point mesh.
        Hollow and filled markers correspond to results without and with the MP2 composite correction (\ref{eq:mp2_comp_corr}), respectively.
        The solid horizontal line represents canonical CCSD reference.
        (b) Same plot as \cref{fig:diamond_eos_conv}a for lithium with LNO-CCSD.
        RPA is used instead of MP2 to correct for the finite-size error and the basis set incompleteness error.
        (c,d) Same plots as \cref{fig:diamond_a0b0_conv}c and d for lithium with LNO-CCSD.
        The solid horizontal line represents canonical CCSD reference.
        The shaded area indicates errors less than $\pm~0.005$~\AA{} in $a_0$ and $\pm~0.5$~GPa in $B_0$.
        }
        \label{fig:lithium_all}
    \end{figure}

    There are two equivalent lithium atoms per unit cell, each having two IAOs of $1s$ and $2s$ character, as displayed in \cref{fig:lithium_orbs}a for a $3\times3\times3$ supercell.
    The associated virtual LNO density [\cref{eq:lno_charge_density}] for LNO-CC and CBNO-CC are displayed in \cref{fig:lithium_orbs}b and c.
    In both theories, the virtual LNOs are seen to localize around the IAOs, with those for CBNO-CC exhibiting more delocalized character, similar to the trend observed in diamond.
    \Cref{fig:lithium_all}a presents the correlation energy of LNO-CCSD and CBNO-CCSD as a function of active space size, evaluated using a TZ basis set with the GTH pseudopotential and a $3\times3\times3$ $k$-point mesh.
    In the absence of the MP2 correction, the LNO-CCSD correlation energy converges quickly to the canonical reference, achieving $1$~kcal/mol of accuracy using only $200$ active orbitals.
    Slower convergence is observed for CBNO-CCSD, which requires approximately $300$ active orbitals to attain the same accuracy.
    This reflects the more delocalized active orbitals in CBNO-CCSD.
    pplying the MP2 composite correction significantly reduces the error of LNO-CCSD for small active spaces of about $100$ orbitals, but the error is similar when more active orbitals are used (similar in magnitude but opposite in sign).
    In contrast, the convergence of CBNO-CCSD is actually worsened by the MP2 correction.

    In \cref{fig:lithium_all}b, we present the counterpoise-correct EOS curves for BCC lithium obtained using LNO-CCSD with different active space size.
    The finite-size errors arising from using a $3\times3\times3$ $k$-point mesh and the basis set incompleteness errors due to using a TZ basis set are corrected approximately with direct random phase approximation (dRPA) (section~S3.1).
    Similar to the case of diamond, the LNO-CCSD energy curves for BCC lithium are smooth for all active space size tested and converge quickly to the canonical CCSD reference.
    An accuracy better than $0.1$~eV in the equilibrium cohesive energy is already achieved with only $123$ active orbitals.
    We then fit the LNO-CCSD energies to the Birch-Murnaghan EOS to extract $a_0$ and $B_0$ for BCC lithium.
    The results are presented in \cref{fig:lithium_all}c and d as a function of active space size.
    The convergence of $a_0$ predicted by LNO-CCSD to the canonical reference is fast and monotonic, with a small error of $\pm~0.005$~\AA{} achieved using approximately $300$ active orbitals.
    The convergence of $B_0$ is even faster: a remarkable accuracy better than $\pm~0.1$~GPa is already achieved with the smallest active space consisting of about $100$ orbitals.
    Including the MP2 composite correction either slightly accelerates the convergence for $a_0$ or maintains the already rapid convergence for $B_0$.

    \begin{table}[!h]
        \centering
        \caption{Thermochemical properties of BCC lithium predicted by different levels of wavefunction theories compared to experiments.}
        \label{tab:lithium_prop}
        \begin{tabular}{lcccl}
            \hline\hline
            Method & $E_{\textrm{coh}}/\textrm{eV}$ & $a_0/\textrm{\AA}$ & $B_0/\textrm{GPa}$ \\
            \hline
            HF & $0.63$ & $3.67$ & $9.6$ & \multirow{2}{*}{this work} \\
            CCSD & $1.43$ & $3.48$ & $13.1$ &  \\
            \hline
            HF & $0.60$ & $3.68$ & $9.0$ & \multirow{2}{*}{ref~\citenum{Neufeld22JPCL}} \\
            CCSD & $1.39$ & $3.49$ & $12.8$ &  \\
            \hline
            HF & $0.56$ & & & \multirow{2}{*}{ref~\citenum{Masios23PRL}}  \\
            CCSD & $1.39$ & & & \\
            \hline
            Experiment & $1.66$ & $3.45$ & $13.3$ & ref~\citenum{Schimka11JCP} \\
            \hline
        \end{tabular}
    \end{table}

    Our final CCSD numbers for the equilibrium cohesive energy, lattice constant, and bulk modulus of BCC lithium are summarized in \cref{tab:lithium_prop}, along with results reported in ref~\citenum{Neufeld22JPCL} and ref~\citenum{Masios23PRL} for comparison.
    From the table, we observe that our CCSD predicted $E_{\textrm{coh}}$, $a_0$, and $B_0$ are are in close agreement with the results from both ref~\citenum{Neufeld22JPCL} and ref~\citenum{Masios23PRL}, with deviations of $0.04$~eV, $-0.01$~\AA{}, and $-0.3$~GPa, which may be due to the use of slightly different basis sets and pseudopotentials.
    The agreement between CCSD and experimental results is not entirely satisfactory, particularly for the cohesive energy.
    This discrepancy highlights the importance of connected triple excitations~\cite{Neufeld23PRL,Masios23PRL}, which are not included at the CCSD level.

    \begin{figure}
        \centering
        \includegraphics[width=5cm]{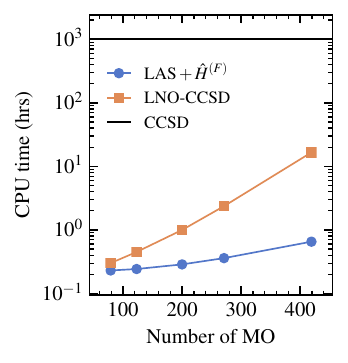}
        \caption{Computational cost of LNO-CCSD as a function of active space size for BCC lithium at equilibrium geometry using a TZ basis set and a $3\times3\times3$ $k$-point mesh.
        The component cost for constructing the local active space and the local Hamiltonian is shown separately in blue.
        The cost of canonical $k$-point CCSD is indicated by the black horizontal line.
        }
        \label{fig:lithium_time}
    \end{figure}

    In \cref{fig:lithium_time}, we present the computational cost of LNO-CCSD for BCC lithium using a TZ basis set and a $3\times3\times3$ $k$-point mesh.
    For active spaces of $271$ and $419$ orbitals, which achieve quantitative accuracy in the thermochemical properties as discussed earlier, LNO-CCSD provides a speedup of approximately $400$-fold and $60$-fold over canonical $k$-point CCSD, respectively.
    For these active spaces, the fragment CCSD calculations dominate the computational cost of LNO-CCSD, as evidenced by the cost for constructing $\mathcal{A}_F$ and $\hat{H}^{(F)}$, also shown in \cref{fig:lithium_time}.

    \section{Conclusion}
    \label{sec:conclusion}

    We have demonstrated the success of local correlation approximations in lowering the cost of periodic CC theory calculations.
    The theory presented in this work can be extended in several ways.
    First, the construction of $\mathcal{A}_F$ and $\hat{H}^{(F)}$ currently has an unfavorable scaling with the supercell size, which will eventually dominate the computational cost for large $N_k$, as needed to eliminate finite-size effects.
    Future work will explore linear-scaling techniques based on local domains and local density fitting~\cite{Nagy16JCTC}, which have already proven successful for molecular LNO-CC~\cite{Nagy19JCTC}.
    Second, for metallic systems, the use of MP2 theory for construction of LNOs and composite corrections to the energy is questionable, given its breakdown in the thermodynamic limit.
    In such cases, we expect that the accuracy of LNO-CC could be further improved by replacing MP2 with a valid but affordable method such as RPA.
    Lastly, although we used CC for solving the local Hamiltonian in this work, the fragment-based protocol presented here is compatible with many other correlated wavefunction methods, providing a generic computational framework for systematically exploiting these methods in materials simulations.

    \section*{Conflict of interest}
    The authors declare no competing conflicts of interest.

    \section*{Acknowledgments}

    This work was supported by the US Air Force Office of Scientific Research under Grant No.~FA9550-21-1-0400 and the Columbia Center for Computational Electrochemistry.
    We acknowledge computing resources from the Flatiron Institute and from Columbia University's Shared Research Computing Facility project, which is supported by NIH Research Facility Improvement Grant 1G20RR030893-01, and associated funds from the New York
    State Empire State Development, Division of Science Technology and Innovation (NYSTAR) Contract C090171, both awarded April 15, 2010.
    The Flatiron Institute is a division of the Simons Foundation.

    \bibliography{refs}

\end{document}


\maketitle

    \vspace{3em}

    \noindent
    Note:
    \begin{enumerate}
        \item figures and equations appearing in the main text will be referred to as ``Fig.~Mxxx'' and ``Eq.~Mxxx'' in this Supplementary Material document.
        \item Further supplementary data can be found in the following Github repository
        \begin{center}
            \url{https://github.com/hongzhouye/supporting_data/tree/main}
        \end{center}
    \end{enumerate}

    \clearpage

    \tableofcontents

    \clearpage

    \section{Expressions for LNO-(T) intermediates}
    \label{sec:lno_t_intermediates}

    The intermediate tensors in the LNO-(T) correlation energy in equation~\fakeref{M16} are defined as (following Nagy and K\'{a}llay, \textit{J.~Chem.~Phys.}~\textbf{146}, 214106, 2017)
    \begin{equation}
    \begin{split}
        &w_{ijk}^{abc} = W_{ijk}^{abc} / \sqrt{-\Delta_{ijk}^{abc}},    \\
        &v_{ijk}^{abc} = V_{ijk}^{abc} / \sqrt{-\Delta_{ijk}^{abc}},    \\
    \end{split}
    \end{equation}
    where
    \begin{equation}
    \begin{split}
        &W_{ijk}^{abc}
            = P_{ijk}^{abc} \left\{
                \sum_{d} V_{bdai} t_{kcjd} -
                \sum_{l} V_{ckjl} t_{ialb}
            \right\}  \\
        &V_{ijk}^{abc}
            = W_{ijk}^{abc} + \frac{1}{2} P_{ijk}^{abc}
                \left\{ V_{aibj} t_{kc} \right\} \\
        &\Delta_{ijk}^{abc}
            = f_{ii} + f_{jj} + f_{kk} - f_{aa} - f_{bb} - f_{cc}
    \end{split}
    \end{equation}
    and $P_{ijk}^{abc}$ generates the following six permutations for a three-particle, three-hole tensor
    \begin{equation}
        P_{ijk}^{abc} \{X_{ijk}^{abc}\}
            = X_{ijk}^{abc} + X_{ikj}^{acb} + X_{jik}^{bac} + X_{jki}^{bca} + X_{kij}^{cab} + X_{kji}^{cba}
    \end{equation}

    \section{Diamond}

    \subsection{Correction for the finite-size and basis set incompleteness errors}

\begin{figure}
        \centering
        \includegraphics[width=0.9\linewidth]{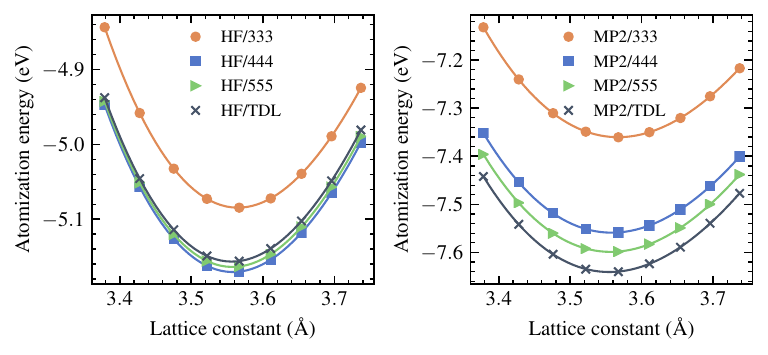}
        \caption{HF (left) and MP2 (right) EOS for different size of $k$-point mesh for diamond.
        The TDL extrapolation is performed using \cref{eq:tdl_extrap} with $4\times4\times4$ and $5\times5\times5$ meshes.
        The markers denote the data from calculations while the lines denote Birch-Murnaghan fit using all data shown.
        All calculations are done using the GTH-cc-pVTZ basis set and the GTH pseudopotential.}
        \label{fig:diamond_hfmp2_eos}
    \end{figure}

    The final equation-of-state (EOS) data presented in fig.~\fakeref{M4} using canonical CCSD and LNO-CCSD/CCSD(T) are obtained as follows
    \begin{equation}    \label{eq:diamond_cc_tdl_cbs_est}
    \begin{split}
        E_{\textrm{coh}}^{\textrm{CC}}(a;\textrm{CBS};\textrm{TDL})
            &\approx E_{\textrm{coh}}^{\textrm{CC}}(a;\textrm{TZ};3\times3\times3)   \\
            &\quad{} + E_{\textrm{coh}}^{\textrm{MP2}}(a;\textrm{TZ};\textrm{TDL})
            - E_{\textrm{coh}}^{\textrm{MP2}}(a;\textrm{TZ};3\times3\times3)    \\
            &\quad{} + E_{\textrm{coh}}^{\textrm{MP2}}(a_0;\textrm{CBS};3\times3\times3)
            - E_{\textrm{coh}}^{\textrm{MP2}}(a_0;\textrm{TZ};3\times3\times3)
    \end{split}
    \end{equation}
    where the CC cohesive energies calculated using a TZ basis set with the GTH pseudopotential and a $3\times3\times3$ $k$-point mesh are corrected by MP2 to the TDL and the CBS limit.
    For LNO-CC, the LNO-CC cohesive energy includes the MP2 composite correction.
    The TDL correction [second line in \cref{eq:diamond_cc_tdl_cbs_est}] is evaluated for each lattice constant using the difference between two MP2 calculations at TZ level, one with the $3\times3\times3$ $k$-point mesh and the other extrapolated to the TDL using a two-point formula
    \begin{equation}    \label{eq:tdl_extrap}
        E(\textrm{TDL})
            = \frac{N_{\bm{k}_1}E(N_{\bm{k}_1}) - N_{\bm{k}_2}E(N_{\bm{k}_2})}
            {N_{\bm{k}_1} - N_{\bm{k}_2}}
    \end{equation}
    where $N_{\bm{k}_1} = 4\times4\times4$ and $N_{\bm{k}_2} = 5\times5\times5$.
    The convergence of the MP2 EOS and the associated bulk properties to the TDL is presented in \cref{fig:diamond_hfmp2_eos} and \cref{tab:diamond_hfmp2_eos}.
    The CBS correction [third line in \cref{eq:diamond_cc_tdl_cbs_est}] is evaluated for the experimental lattice constant ($a_0 = 3.567~\mathrm{\AA}$) only using the difference between two MP2 calculations using the $3\times3\times3$ $k$-point mesh, one with the TZ basis set and the other extrapolated to the CBS limit using a two-point formula
    \begin{equation}    \label{eq:cbs_extrap}
        E(\textrm{CBS})
            = \frac{X_1^3 E(X_1) - X_2^3 E(X_2)}
            {X_1^3 - X_2^3}
    \end{equation}
    where $X_1 = 3$ (TZ) and $X_2 = 4$ (QZ).
    The resulting value is a rigid shift of ca.~$-0.23$~eV for all EOS curves reported in fig.~\fakeref{M4}.

    \begin{table}
        \caption{Convergence of HF and MP2 bulk properties of diamond to the TDL with the GTH-cc-pVTZ basis set and the GTH pseudopotential.
        A composite correction accounting for the basis set incompleteness error is used to correct the cohesive energy to the CBS limit as discussed in the text.}
        \label{tab:diamond_hfmp2_eos}
        \begin{tabular}{ccccccc}
            \hline\hline
            \multirow{2}*{$k$-mesh} & \multicolumn{3}{c}{HF} & \multicolumn{3}{c}{MP2}  \\
            \cmidrule(lr){2-4} \cmidrule(lr){5-7}
            & $E_{\textrm{coh}}/$eV & $a_0/\mathrm{\AA}$ & $B_0/$GPa
            & $E_{\textrm{coh}}/$eV & $a_0/\mathrm{\AA}$ & $B_0/$GPa  \\
            \hline
            $3\times3\times3$ & $5.08$ & $3.565$ & $487.2$ & $7.36$ & $3.567$ & $446.6$   \\
            $4\times4\times4$ & $5.17$ & $3.558$ & $489.9$ & $7.56$ & $3.558$ & $448.9$   \\
            $5\times5\times5$ & $5.16$ & $3.557$ & $490.0$ & $7.60$ & $3.556$ & $449.0$   \\
            \hline
            TDL   & $5.16$ & $3.557$ & $490.0$ & $7.64$ & $3.554$ & $449.2$   \\
            \hline
            $\Delta_{\textrm{CBS}}$ & $0.01$ & & & $0.23$ & & \\
            TDL/CBS & $5.17$ & & & $7.87$ & & \\
            \hline
        \end{tabular}
    \end{table}

    \subsection{Impact of LNO choices}

\begin{figure}[!h]
        \centering
        \includegraphics[width=15cm]{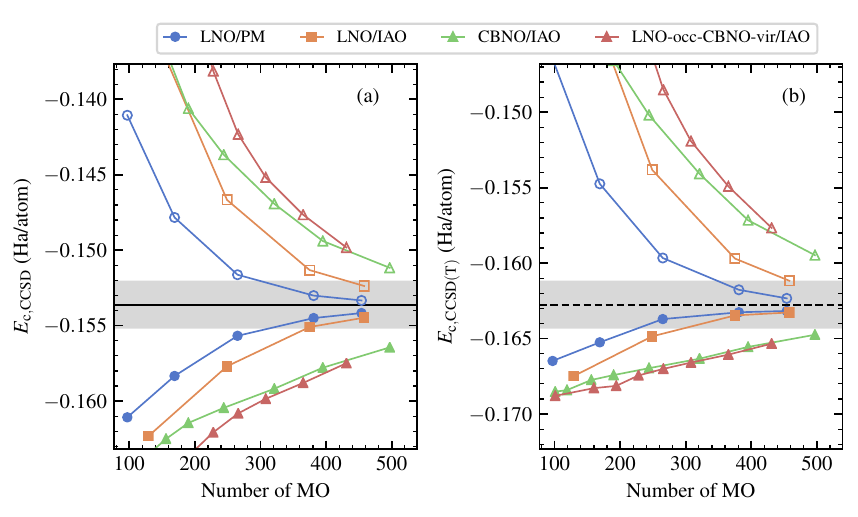}
        \caption{Same plot as fig.~\fakeref{M3} except for the inclusion of results obtained using a hybrid choice of LNOs (LNO-occ-CBNO-vir) as described in the main text.}
        \label{fig:diamond_lno_cmp}
    \end{figure}

    \Cref{fig:diamond_lno_cmp} is the same as fig.~\fakeref{M3} except for the inclusion of results obtained using a hybrid choice of occupied LNOs from LNO-CC and virtual LNOs from CBNO-CC (with a fixed $\eta_{\textrm{occ}} / \eta_{\textrm{vir}} = 10$).
    The LNO-occ-CBNO-vir-CC shows similar slow convergence compared to CBNO-CC, especially for large active space.
    As discussed in the main text, this suggests that the virtual LNOs in CBNO-CC are responsible for the slow convergence of the CBNO-CC correlation energy with the active space size.

    \subsection{EOS from LNO-CC/IAO}

\begin{figure}[!h]
        \centering
        \includegraphics[width=12cm]{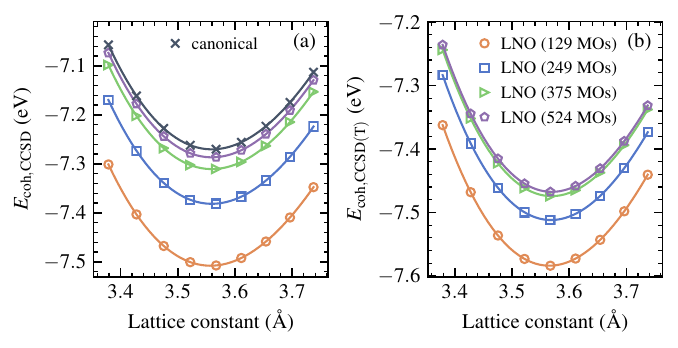}
        \caption{Same plot as fig.~\fakeref{M4} except that IAOs are used instead of PM orbitals as the LOs.}
        \label{fig:diamond_lno_eos_iao}
    \end{figure}

    The LNO-CC EOS results obtained using IAOs as the LOs are shown in \cref{fig:diamond_lno_eos_iao}.
    Compared to the results in fig.~\fakeref{M4} obtained using PM orbitals, the use of IAOs leads to overall similar convergence pattern but requires about $100$ more active orbitals to achieve a similar accuracy.
    This is consistent with the convergence results shown in fig.~\fakeref{M3}.

    \subsection{Convergence of all-electron LNO-CC calculations}

\begin{figure}[!h]
        \centering
        \includegraphics[width=12cm]{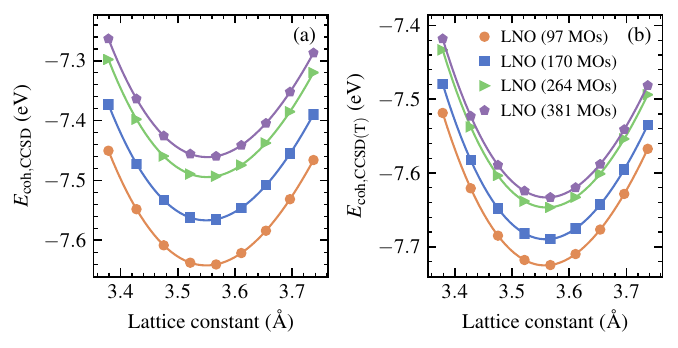}
        \caption{Same plot as fig.~\fakeref{M4} except that all calculations were performed using the all-electron potential with Dunning's cc-pV$X$Z basis sets.}
        \label{fig:diamond_alle_eos}
    \end{figure}

\begin{figure}[!h]
        \centering
        \includegraphics[width=12cm]{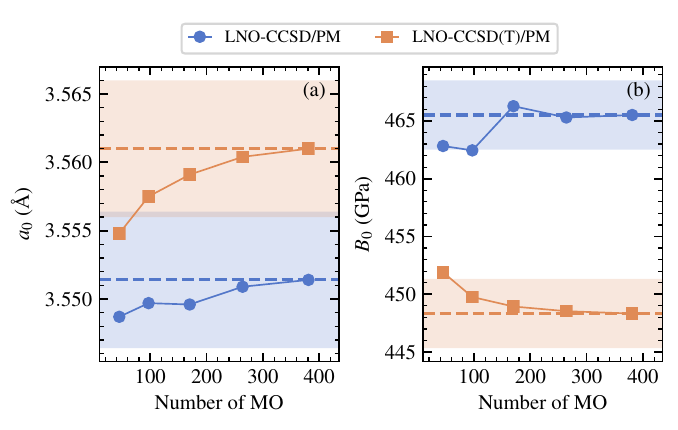}
        \caption{Same plot as fig.~\fakeref{M5} except that all calculations were performed using the all-electron potential with Dunning's cc-pV$X$Z basis sets.}
        \label{fig:diamond_alle_a0b0}
    \end{figure}

    The all-electron LNO-CC calculations of the EOS and the resulting convergence of $a_0$ and $B_0$ are shown in \cref{fig:diamond_alle_eos,fig:diamond_alle_a0b0}.
    The $1s$ core electrons are kept frozen during the correlated calculations.
    These results follow closely the pseudopotential-based results shown in figs.~\fakeref{M4} and \fakeref{M5}, respectively.

    \begin{table}
        \caption{Same data as in \cref{tab:diamond_hfmp2_eos} except using all-electron potential and Dunning's cc-pV$X$Z basis sets.}
        \label{tab:diamond_alle_hfmp2_eos}
        \begin{tabular}{ccccccc}
            \hline\hline
            \multirow{2}*{$k$-mesh} & \multicolumn{3}{c}{HF} & \multicolumn{3}{c}{MP2}  \\
            \cmidrule(lr){2-4} \cmidrule(lr){5-7}
            & $E_{\textrm{coh}}/$eV & $a_0/\mathrm{\AA}$ & $B_0/$GPa
            & $E_{\textrm{coh}}/$eV & $a_0/\mathrm{\AA}$ & $B_0/$GPa  \\
            \hline
            $3\times3\times3$ & $5.24$ & $3.558$ & $495.4$ & $7.54$ & $3.559$ & $454.0$ \\
            $4\times4\times4$ & $5.31$ & $3.552$ & $498.1$ & $7.73$ & $3.550$ & $455.7$ \\
            $5\times5\times5$ & $5.30$ & $3.551$ & $498.1$ & $7.76$ & $3.549$ & $456.8$ \\
            \hline
            TDL   & $5.30$ & $3.550$ & $498.2$ & $7.80$ & $3.547$ & $458.1$ \\
            \hline
            $\Delta_{\textrm{CBS}}$ & $0.01$ & & & $0.23$ & & \\
            TDL/CBS & $5.31$ & & & $8.04$ & & \\
            \hline
        \end{tabular}
    \end{table}

    \section{BCC lithium}

    \subsection{Correction for the finite-size and basis set incompleteness errors}

\begin{figure}[!h]
        \centering
        \includegraphics[width=12cm]{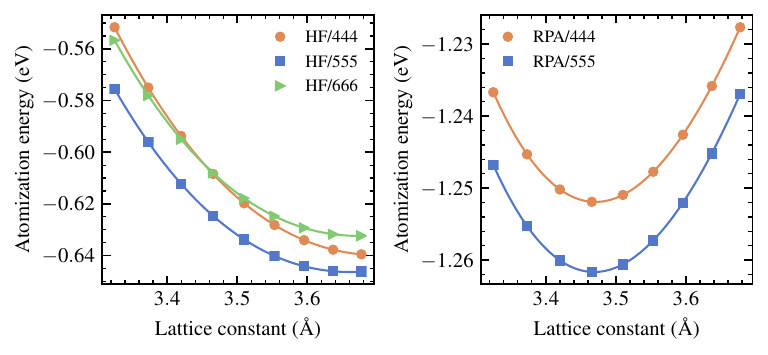}
        \caption{EOS of BCC lithium calculated by HF (left) and RPA (right) with different Brillouin zone sampling.
        The RPA total energy of a given $k$-point mesh size is obtained by adding the RPA correlation energy evaluated using the corresponding $k$-point mesh to the HF energy obtained using the largest $k$-point mesh, $6\times6\times6$.
        }
        \label{fig:lithium_hfrpa_eos}
    \end{figure}

    The final equation-of-state (EOS) data presented in fig.~\fakeref{M8b} using canonical CCSD and LNO-CCSD are obtained as follows
    \begin{equation}    \label{eq:lithium_cc_tdl_cbs_est}
    \begin{split}
        E_{\textrm{coh}}^{\textrm{CC}}(a;\textrm{CBS};\textrm{TDL})
            &\approx E_{\textrm{coh}}^{\textrm{CC}}(a;\textrm{TZ};3\times3\times3)   \\
            &\quad{} + E_{\textrm{coh}}^{\textrm{RPA}}(a;\textrm{TZ};\textrm{TDL})
            - E_{\textrm{coh}}^{\textrm{RPA}}(a;\textrm{TZ};3\times3\times3)    \\
            &\quad{} + E_{\textrm{coh}}^{\textrm{RPA}}(a_0;\textrm{CBS};3\times3\times3)
            - E_{\textrm{coh}}^{\textrm{RPA}}(a_0;\textrm{TZ};3\times3\times3)
    \end{split}
    \end{equation}
    where the CC cohesive energies calculated using a TZ basis set with the GTH pseudopotential and a $3\times3\times3$ $k$-point mesh are corrected by RPA to the TDL and the CBS limit.
    For LNO-CC, the MP2 composite correction is not included.
    The TDL correction [second line in \cref{eq:lithium_cc_tdl_cbs_est}] is evaluated for each lattice constant using the difference between two RPA calculations at TZ level, one with the $3\times3\times3$ $k$-point mesh and the other with a $5\times5\times5$ $k$-point mesh.
    The latter is taken approximately as the TDL results, as justified by the convergence of the RPA EOS curve and the associated bulk properties with $k$-point mesh size as shown in \cref{fig:lithium_hfrpa_eos} and \cref{tab:lithium_hfrpa_eos}.
    The CBS correction [third line in \cref{eq:lithium_cc_tdl_cbs_est}] is evaluated for the experimental lattice constant ($a_0 = 3.567~\mathrm{\AA}$) only using the difference between two RPA calculations using the $3\times3\times3$ $k$-point mesh, one with the TZ basis set and the other extrapolated to the CBS limit using the two-point formula (\ref{eq:cbs_extrap}) with $X_1 = 3$ (TZ) and $X_2 = 4$ (QZ).
    The resulting value is a rigid shift of ca.~$+0.05$~eV for all EOS curves reported in fig.~\fakeref{M8b}.

    \begin{table}[!h]
        \caption{Convergence of HF and MP2 bulk properties of BCC lithium with respect to Brillouin sampling using the GTH-cc-pVTZ basis set and the GTH pseudopotential.
        All RPA results are obtained by adding the RPA correlation energy evaluated using the corresponding $k$-point mesh to the HF energy obtained using the $6\times6\times6$ $k$-point mesh.
        The HF/$6\times6\times6$ and RPA/$5\times5\times5$ are taken as approximate TDL results.
        The cohesive energy is further corrected by a composite correction to the CBS limit as discussed in the text.}
        \label{tab:lithium_hfrpa_eos}
        \begin{tabular}{ccccccc}
            \hline\hline
            \multirow{2}*{$k$-mesh} & \multicolumn{3}{c}{HF} & \multicolumn{3}{c}{MP2}  \\
            \cmidrule(lr){2-4} \cmidrule(lr){5-7}
            & $E_{\textrm{coh}}/$eV & $a_0/\mathrm{\AA}$ & $B_0/$GPa
            & $E_{\textrm{coh}}/$eV & $a_0/\mathrm{\AA}$ & $B_0/$GPa  \\
            \hline
            $3\times3\times3$ & $0.96$ & $3.574$ & $11.4$ & $1.10$ & $3.501$ & $12.7$ \\
            $4\times4\times4$ & $0.64$ & $3.701$ & $9.2$ & $1.25$ & $3.471$ & $13.3$  \\
            $5\times5\times5$ & $0.65$ & $3.661$ & $9.7$ & $1.26$ & $3.469$ & $13.4$  \\
            $6\times6\times6$ & $0.63$ & $3.673$ & $9.6$ & & & \\
            \hline
            Approx.~TDL & $0.63$ & $3.673$ & $9.6$ & $1.26$ & $3.469$ & $13.4$    \\
            \hline
            $\Delta_{\textrm{CBS}}$ & $0.00$ & & $-0.05$ & & \\
            Approx.~TDL/CBS & $0.63$ & & $1.21$ & & \\
            \hline
        \end{tabular}
    \end{table}
